\documentclass[journal]{IEEEtran}
\pdfoutput=1
\usepackage{amsmath} % Allows the mathematical features found in AMS
\usepackage[english]{babel}

\usepackage{graphicx} % Required for including pictures
\usepackage{enumerate}
\usepackage{graphics}
\usepackage{enumitem}
\usepackage{amssymb}
\usepackage{float}
\usepackage{rotating}
\usepackage{caption}
\usepackage{subcaption}
\usepackage{copyrightbox}
\usepackage{rotating}
\usepackage{booktabs}
\usepackage{colortbl}
\usepackage{multicol}
\usepackage{multirow}

\usepackage{cite}
\usepackage{hyperref}

\ifCLASSINFOpdf
\else
\fi
\newcommand{\proba}{\text{p}}
\DeclareMathOperator*{\argmin}{argmin}
\DeclareMathOperator{\E}{\mathbb{E}}
\newcommand{\V}[1]{\boldsymbol{#1}}

\newcommand{\speckle}{{S}}
\newcommand{\reflectivity}{{X}}
\newcommand{\intensity}{{Y}}

\newcommand{\logreflectivity}{\logtransformed{{X}}}
\newcommand{\logintensity}{\logtransformed{{Y}}}
\newcommand{\logspeckle}{\logtransformed{{S}}}

\renewcommand{\logreflectivity}{{x}}
\renewcommand{\logintensity}{{y}}
\renewcommand{\logspeckle}{{s}}

\newcommand{\FIXME}[1]{\textcolor{red}{#1}}

\renewcommand{\speckle}{{u}}
\renewcommand{\reflectivity}{{v}}
\renewcommand{\intensity}{{w}}

\begin{document}
\title{SAR2SAR: a semi-supervised despeckling algorithm for SAR images}

\author{Emanuele~Dalsasso, Lo{\"i}c~Denis, Florence~Tupin%
\thanks{E. Dalsasso and F. Tupin are with LTCI, Télécom Paris, Institut Polytechnique de Paris, Palaiseau, France, e-mail: forename.name@telecom-paris.fr.}%
\thanks{L. Denis is the Univ Lyon, UJM-Saint-Etienne, CNRS, Institut d Optique Graduate School, Laboratoire Hubert Curien UMR 5516, F-42023, SAINT-ETIENNE, France, e-mail: loic.denis@univ-st-etienne.fr.}%
}
\maketitle
\begin{abstract}
Speckle reduction is a key step in many remote sensing applications. By strongly affecting synthetic aperture radar (SAR) images, it makes them difficult to analyse. Due to the difficulty to model the spatial correlation of speckle, a deep learning algorithm with semi-supervision is proposed in this paper: SAR2SAR. Multi-temporal time series are leveraged and the neural network learns to restore SAR images by only looking at noisy acquisitions. To this purpose, the recently proposed noise2noise framework \cite{lehtinen2018noise2noise} has been employed. The strategy to adapt it to SAR despeckling is presented, 
%through the use of a priori information to take into account changes
based on a compensation of temporal changes
and a loss function adapted to the statistics of speckle.

A study with synthetic speckle noise is presented to compare the performances of the proposed method with other state-of-the-art filters. Then, results on real images are discussed, to show the potential of the proposed algorithm. The code is made available to allow testing and reproducible research in this field.

\end{abstract}

\IEEEpeerreviewmaketitle

\section{Introduction}
\IEEEPARstart{S}{ynthetic} Aperture Radar (SAR)
is an active imaging technology that is widely used for Earth observation, thanks to its capability of acquiring images 
%all-day 
by day or night
and in (almost) all-weather conditions. Agriculture, forestry, oceanography are among the fields that benefit from the exploitation of SAR images, employed in a wide range of practical applications such as urban monitoring, land-use mapping, biomass estimation, damage assessment, oil spill detection, ice monitoring, among others \cite{moreira2013tutorial}. This is achieved through advanced techniques, like interferometry and polarimetry.

However, interpreting SAR images is a challenging task, both for human observers and for automatic tools aiming at extracting useful information. Indeed, they are corrupted by \textit{speckle}. Although commonly referred to as noise (we will also adopt this convention in this paper), speckle is a physical phenomenon that is caused by the coherent sum of the contributions from different elementary scatterers within the same resolution cell, which the radar cannot resolve. The phase differences induce fluctuations in the complex summation and then in the observed amplitude, that produces in the observed image a granular behaviour.
%, or "salt and pepper" noise.

In this context, being capable of effectively removing speckle from SAR images is of crucial importance for the community. Great efforts have been devoted to this topic. Among the most sophisticated techniques recently developed, one can mention non-local (NL) algorithms \cite{deledalle2009iterative, deledalle2010nl, deledalle2010polarimetric, parrilli2012nonlocal}, generalized in NL-SAR \cite{deledalle2014nl}, a fully automatic algorithm that handles any SAR modality, single- or multi-look images, by performing several non-local estimations to best restore speckle-free data.  

% FT ? A general framework that allows to exploit any Gaussian denoiser in its iterative speckle removal process is MuLoG. Recent advances of deep learning for additive white Gaussian noise (AWGN) removal can also be exploited 
%motivates the spark of interest for it in the radar community. In \cite{deledalle2017mulog}, Deledalle \textit{et al.} propose a general framework that allows to employ any Gaussian denoiser in its iterative speckle removal process: results using a convolutional neural network (CNN) are proposed in \cite{yang2019sar}.

In \cite{deledalle2017mulog}, a general framework, called MuLoG, is proposed to apply any image denoiser originally designed for additive Gaussian noise within an iterative speckle removal procedure. Recent advances of deep learning for additive white Gaussian noise (AWGN) removal can thus be exploited \cite{yang2019sar}.
Training an end-to-end model for speckle reduction has been most recently studied \cite{chierchia2017sar, wang2017sar, wang2017generative, zhang2018learning,lattari2019deep, molini2020towards}. However, taking into account the peculiarities of SAR data remains challenging: noise distribution is not the same as in natural images, as well as the content, the texture, or the physical meaning of a pixel value. On top of that, there is an inherent scarcity of speckle-free references to train supervised deep learning algorithms to map a noisy SAR image to a speckle-free image. Thus, borrowing algorithms proposed in the computer vision field and extending them to the speckle removal task is not straightforward. 

In this paper, we address the lack of noise-free references by extending the \texttt{noise2noise} approach proposed by Lehtinen \textit{et al.} in \cite{lehtinen2018noise2noise} in order to take into account the peculiarities of SAR data. Thanks to a grounded loss function formulation, which depends on the speckle model, large stacks of multitemporal images acquired over the same area are exploited in two ways. In the first instance, as described in \cite{dalsasso2020sar}, a dataset of noiseless images is created and our model is trained with synthetic speckle, following the fully developed Goodman’s model of noise \cite{goodman1976some}. At a later stage, we feed the network with real acquisitions, allowing learning of the spatial correlation introduced by the SAR processing steps, namely spectral windowing and oversampling \cite{abergel2018subpixellic} \cite{dalsasso2020handle}. The problem of temporal changes is addressed by a strategy for change compensation.
%thanks to the injection of \textit{a priori} information.
This ensures robustness and generalization of the algorithm, since any temporal series of SAR images can be used at this transfer learning step.

\section{Related works}

SAR images lack of noise-free references, which makes it non-trivial to adapt deep learning image denoising methods to speckle reduction. An analysis of the solutions recently proposed is carried out in this section.

The first paper that investigates the use of CNN for SAR image despeckling has been proposed by Chierchia \textit{et al.} \cite{chierchia2017sar}. Inspired by Zhang \textit{et al.} \cite{zhang2017beyond}, their SAR-CNN is an adaptation of the denoising CNN (DnCNN) to SAR images. The groundtruth is created exploiting temporal series of images: assuming that no change has occurred between acquisitions, the images are temporally multilooked to produce a reference image. 
%The network is then trained in a supervised way. A residual learning framework is adopted: by employing the homomorphic transform (i.e. data are log-transformed to achieve variance stabilization \cite{deledalle2017mulog}), the networks learns to reproduce the residual speckle noise, which is then subtracted from the input image to produce an estimate of the speckle-free image. 
While achieving high-quality results both on images with synthetic noise and on real SAR images, the method is difficult to reproduce. Not only is it rare to observe temporal stability, but the definition of absence of change is ambiguous. Only images with short temporal baseline generally offer sufficient temporal stability, but this often comes with a strong temporal correlation of speckle, which undermines the ability of the network to efficiently remove speckle. The same analysis bears for \cite{cozzolino2020nonlocal}, where a non-local CNN is instead trained.

An alternative approach considered by Wang \textit{et al.} \cite{wang2017sar,wang2017generative} and by Zhang \textit{et al.} \cite{zhang2018learning} consists in using natural images and produces SAR-like data by generating synthetic speckle, following Goodman's model \cite{goodman1976some}. Only the case $L=4$ of multilooked images has been considered in the experiments, 
%However, it appears that the L=4 case has been studied, 
making the speckle fluctuations less prominent than in the most interesting case of single-look SAR acquisitions.

%Results with synthetic 4-look speckle noise are also proposed by Zhang \textit{et al.} \cite{zhang2018learning}. The dataset is composed of natural images. Dilated (or \emph{à trous}) convolutions consist in sparsifying the convolutional kernel to enlarge the receptive field, while maintaining a low number of parameters: a light architecture, less prone to overfitting, can thus be obtained. They are combined with a residual learning strategy in the so called SAR dilated residual network (SAR-DRN)

The use of natural images, combined with synthetic speckle noise, has the advantage that a huge dataset can be effortlessly generated, allowing the training of models with numerous parameters (i.e., deep architectures). However, peculiar characteristics of SAR images are neglected: content, geometry, resolution, scattering phenomena, etc. A compromise is proposed by Lattari \textit{et al.} \cite{lattari2019deep}. While the network is initially trained on a synthetic dataset built from natural images, a fine-tuning on SAR images is subsequently carried out. To this purpose, stacks of images are temporally averaged to produce a target image, then corrupted with synthetic speckle. In this way, the model can better handle real SAR images. To this end, a U-Net architecture \cite{ronneberger2015u} is employed in a residual fashion, along with the homomorphic approach. In \cite{shen2020sar}, to overcome the limits of a data-driven approach in modeling the speckle, a deep CNN is integrated with a model-driven method based on the assumption that the speckle follows a Gamma distribution.

At present, there does not exist a clear strategy on how to train deep learning models for SAR image despeckling. Some insight is given in \cite{dalsasso2020sar}, where a high-quality dataset of noise-free SAR images is built to train an end-to-end deep learning model. 

Algorithms developed using speckle generated under Goodman’s fully developed speckle model generally assume an absence of spatial correlations \cite{goodman1976some}, which is not the case in actual SAR images synthetized by space agencies \cite{argenti2013tutorial, abergel2018subpixellic}. Thus, a careful pre-processing step must be performed before handling real images to prevent the apparition of strong artifacts \cite{dalsasso2020handle}. Whitening the spectrum \cite{abergel2018subpixellic, lapini2013blind} or down-sampling the image are possible strategies \cite{dalsasso2020handle}. Yet, they are either not easy to apply in a systematic fashion or they result in a loss of spatial resolution. 

To overcome these issues, an end-to-end self-supervised deep learning model trained on real SAR images is proposed by Boulch \textit{et al.} \cite{boulch2018learning}. Their work is based on the intuition that, if no change occurs, randomly picking two images from a temporal stack
%at each iteration 
and training a neural network to reproduce one image starting from the other eventually leads the network to output the underlying speckle-free reflectivity, as only the speckle realization is changing.

As temporal stability is rarely observed in practise, areas affected by changes can be masked out in the loss \cite{ma2020sar} or independent noisy image pairs can be generated from a noisy image thanks to a generative approach \cite{yuan2019practical}. Alternatively, Molini \textit{et al.} \cite{molini2020speckle2void} present an algorithm enabling direct training on real images, learning to denoise from a single image at a time. Their method however relies on the assumption that speckle is spatially uncorrelated, i.e. a whitening preprocessing stage is shown to be crucial in order to achieve good performances
%Assuming spatially independent pixels, this method relies on a whitening stage, which is crucial in order to achieve good performances.
%The authors suggest further studies on real images, while showing promising preliminary results. %EMANUELE: j'ai modifié la référence et cité Speckle2Void, ce qui rend cette phrase pas pertinente.

\section{Speckle model}\label{sec:specklemodel}
The fluctuations affecting SAR images arise from the 3-D spatial configuration and the nature of the scatterers inside a resolution cell. %Being indistinguishable from the radar, the e
Echoes generated by each scatterer interfere either in a constructive or in a destructive way. 

These perturbations are generally modeled as a multiplicative noise, i.e. the speckle. Assuming a large number of elementary scatterers, producing echoes with independent and identically distributed (i.i.d.) complex amplitudes, the fully-developed speckle model proposed by Goodman \textit{et al.} \cite{goodman1976some} relates the measured intensity $\intensity$, the underlying reflectivity $\reflectivity$, and the speckle $\speckle$ as follows:
\begin{equation}\label{eq:multiplicative_model}
    \intensity=\reflectivity \times \speckle\,,
\end{equation}
where the speckle component is modeled by a random variable distributed according to a gamma distribution:
\begin{equation}\label{eq:gamma_distribution}
    \proba(\speckle) =\frac{L^L}{\Gamma(L)}\speckle^{L-1}\exp\left(-L\speckle \right)\,,
\end{equation}
with $L\geq 1$ the number of looks and $\Gamma(\cdot)$ the gamma function. It follows that $\mathbb{E}[\speckle]=1$ and $\text{Var}[\speckle]=1/L$. Thus, averaging independent samples in intensity leads to an unbiased estimator of the underlying reflectity, reducing the fluctuations by a factor proportional to the number of samples available. 

In order to stabilize the variance, i.e. to make it independent from the reflectivity, 
a logarithmic transformation is often applied (homomorphic transform, see
%it is often convenient to employ the homomorphic transform 
\cite{deledalle2017mulog}). 
%Through a logarithmic transformation, t
The log-speckle $\logspeckle$ has an additive behaviour:
\begin{equation}\label{eq:additive_model}
    \logintensity = \logreflectivity + \logspeckle\,,
\end{equation}
where $\logspeckle$ follows a Fisher-Tippett distribution described by:
\begin{equation}\label{eq:fishertippett_distribution}
    \proba(\logspeckle) = 
    \frac{L^L}{\Gamma(L)}e^{L\logspeckle}\cdot \exp(-Le^{\logspeckle})\,.
\end{equation}
The log-speckle has a variance that does not depend on the log-reflectivity, i.e., it is
stationary throughout the image: $\text{Var}[\logspeckle] = \psi(1,L)$, where $\psi(1, L)$ is the polygamma function of order $L$ \cite{abramowitz1965handbook}. The mean of the log-speckle is not zero: $\mathbb{E} [\logspeckle] = \psi(L)-\log(L)$, with $\psi$ the digamma function. Averaging log-transformed intensities $\logintensity$ requires a compensation for $\psi(L)-\log(L)$ to obtain an unbiased estimator of log-reflectivity.
%and has a non-zero mean, namely $\mathbb{E} [\logspeckle] = \psi(L)-\log(L)$ (where $\psi$ is the digamma function). Thus, averaging log-transformed intensities $\logintensity$ leads to a biased estimator of the true reflectivity. However, we achieve stationarity of the variance over the image, with $\text{Var}[\logspeckle] = \psi(1,L)$ (where $\psi(1, L)$ is the polygamma function of order $L$ \cite{abramowitz1965handbook}).

The focusing of a SAR image involves a series of processing steps that, as a result, introduce a %strong
spatial correlation between neighboring pixels. Goodman's speckle model does not take these correlations into account, requiring an adaptation of the algorithms relying on the fully-developed i.i.d assumptions \cite{lapini2013blind}\cite{dalsasso2020handle}.

\section{Semi-supervised image restoration: from Gaussian denoising to speckle reduction}%Fisher-Tippett distribution}
% LOIC: j'ai raccourci / reformulé le titre de la section, à valider...

\subsection{Self-supervised learning}
In the supervised learning setting, pairs $(\V x,\V y)$ of noiseless and noisy images are available for training. A common approach to estimate the parameters $\V \theta$ of an estimator $f_{\V \theta}:\;\V y\mapsto \hat{\V x}=f_{\V \theta}(\V y)$ is to minimize the $\ell_2$ loss function:
\begin{equation}
    %\hat{x}^{(\ell_2)}=
    \hat{\V \theta}_{\text{supervised}}^{(\ell_2)}\in\argmin_{\V \theta} \;
    %\E_{X}\bigl[\E_{Y|X} \{(f_\theta(y)-x)^2\}\bigr]\,,
    \E_{X} \bigl[\|f_{\V \theta}(\V y)-\V x\|^2\bigr]\,,
    \label{eq:supervl2}
\end{equation}
where $\V x$ is a random realization of the random vector $X$ and $\V y$ is a random realization under the conditional distribution $\proba_{Y|X}$.

The self-supervised approach \texttt{noise2noise} introduced by Lehtinen \textit{et al.} \cite{lehtinen2018noise2noise} considers only \emph{noisy pairs} $(\V y_1,\V y_2)$, where $\V y_1$ and $\V y_2$ are two independent realizations drawn under the same conditional distribution $\proba_{Y|X}$. The authors suggest replacing the unknown realization $\V x$ with the noisy observation $\V y_2$ given that it is much easier to obtain additional noisy measurements of a static scene rather than very high quality measurements (i.e., virtually noise-free images):
\begin{equation}
    %\hat{x}^{(\ell_2)}=
    \hat{\V \theta}_{\text{self-supervised}}^{(\ell_2)}\in\argmin_{\V \theta} \;
    %\E_{X}\bigl[\E_{Y|X} \{(f_\theta(y)-x)^2\}\bigr]\,,
    \E_{X} \bigl[\|f_{\V \theta}(\V y_1)-\V y_2\|^2\bigr]\,.
    \label{eq:unsupervl2}
\end{equation}
The rationale behind this substitution comes from the conditional expectation of the expansions of the squared $\ell_2$ norms in equations (\ref{eq:supervl2}) and (\ref{eq:unsupervl2}):
\begin{align}
\E_{Y|X} \bigl[\|f_{\V \theta}(\V y)-\V x\|^2\bigr]&=\E_{Y|X} \bigl[\|f_{\V \theta}(\V y)\|^2\bigr]\nonumber\\
&\hspace*{-0.5em}-2\V x^{\text{t}}\E_{Y|X} \bigl[f_{\V \theta}(\V y)\bigr]+\|\V x\|^2\,,\\%\nonumber\\
\E_{Y|X} \bigl[\|f_{\V \theta}(\V y_1)-\V y_2\|^2\bigr]&=\E_{Y|X} \bigl[f_{\V \theta}(\V y_1)^2\bigr]\nonumber\\
&\hspace*{-8em}-2\E_{Y|X} \bigl[\V y_2\bigr]^{\text{t}}\E_{Y|X} \bigl[f_{\V \theta}(\V y_1)\bigr]+\E_{Y|X} \bigl[\|\V y_2\|^2\bigr]\,.
\end{align}
%$\E_{Y|X} \bigl[\|f_{\V \theta}(\V y)-\V x\|^2\bigr]=\E_{Y|X} \bigl[\|f_{\V \theta}(\V y)\|^2\bigr]-2\V x^{\text{t}}\E_{Y|X} \bigl[f_{\V \theta}(\V y)\bigr]+\|\V x\|^2$ and $\E_{Y|X} \bigl[\|f_{\V \theta}(\V y_1)-\V y_2\|^2\bigr]=\E_{Y|X} \bigl[f_{\V \theta}(\V y_1)^2\bigr]-2\E_{Y|X} \bigl[\V y_2\bigr]^{\text{t}}\E_{Y|X} \bigl[f_{\V \theta}(\V y_1)\bigr]+\E_{Y|X} \bigl[\|\V y_2\|^2\bigr]$.
Provided that the noise is centered, i.e. $\E_{Y|X} \bigl[\V y_2\bigr]=\V x$, the two expansions differ only by a term that is constant with respect to the parameters $\V \theta$. Therefore, if the training set is large enough, parameters estimated with the self-supervised procedure of (\ref{eq:supervl2}) are equivalent to parameters estimated with the supervised procedure (\ref{eq:unsupervl2}).

\begin{figure}[t]
	\centering
	\includegraphics[width=\columnwidth]{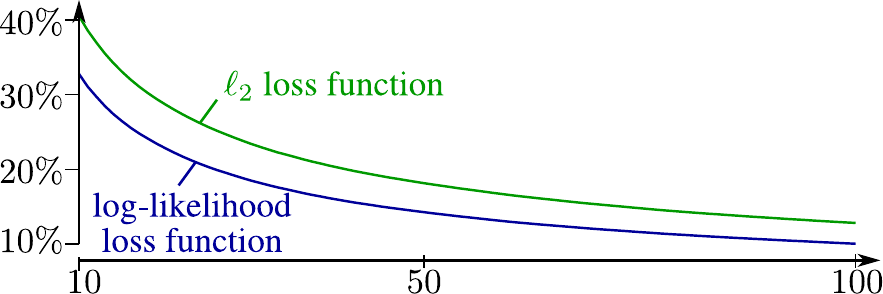}
	\caption{Evolution of the root mean square estimation error as a function of the number of samples.}
	\label{fig:effic}
\end{figure}

In practice, training sets are limited and it is therefore necessary to consider how fast the self-supervised estimator converges to the supervised estimator. Under non-Gaussian noise, other loss functions may be more efficient. This is in particular the case of the co-log-likelihood:
\begin{equation}
    %\hat{x}^{(\ell_2)}=
    \hat{\V \theta}_{\text{self-supervised}}^{(\text{lik})}\in\argmin_{\V \theta} \;
    %\E_{X}\bigl[\E_{Y|X} \{(f_\theta(y)-x)^2\}\bigr]\,,
    \E_{X} \bigl[-\log\proba(\V y_2|f_{\V \theta}(\V y_1))\bigr]\,.
    \label{eq:unsupervllk}
\end{equation}
Among the M-estimators, i.e. methods to estimate parameters $\V \theta$ based on the minimization of a loss function over the training set, the maximum likelihood estimators are known to be efficient \cite{van2004detection}. This is illustrated in the case of speckle in figure \ref{fig:effic}, where the root mean square error\footnote{since the log-speckle is not centered, a compensation is added to prevent a bias with the $\ell_2$ loss, which is replaced by $\|f_{\V \theta}(\V y_1)-\V y_2+\psi(L)-\log L\|^2$} of the log-intensity is reported for the $\ell_2$ and the log-likelihood loss functions. The minimizer of the log-likelihood loss converges more quickly to $\V x$, which indicates that it should be preferred as a loss function for self-supervised training of a despeckling network and is confirmed in our experiments described in section \ref{sec:expe}.
%The efficiency of the maximum likelihood estimator 

\subsection{Application to SAR image despeckling: a semi-supervised approach}
When $\V y_1$ and $\V y_2$ are noisy log-intensity images, section \ref{sec:specklemodel} recalled that the conditional distribution $\proba_{Y|X}(\V y|\V x)$ is a Fisher-Tippett distribution. The loss function in (\ref{eq:unsupervllk}) takes the form, under a simplifying assumption of statistical independence between pixels:
\begin{align}\label{eq:lossSAR}
    \ell\big(f_{\V \theta}(\V y_1),\V y_2\big)& = -\log\proba(\V y_2|f_{\V \theta}(\V y_1))\nonumber\\ &=\sum_k f_{\V \theta}([\V y_1]_k)-[\V y_2]_k\nonumber\\[-2ex]
    &\qquad\qquad\quad
    +\exp\bigl([\V y_2]_k-f_{\V \theta}([\V y_1]_k)\bigr)\,,
    %\big(f_{\text{CNN}}(\logintensity_i)-\logintensity_j\big)^2\,,
\end{align}
where the constant offset and the multiplicative factor $L$ are dropped since they are irrelevant in the minimization problem (\ref{eq:unsupervllk}), and the sum involves all the pixel values $[\V y_1]_k$ and $[\V y_2]_k$ of the image pair.
\label{sec:sar2sar_losses}

Beyond the adaptation of the loss function to the statistics of speckle, it is necessary to account for changes that occur between two SAR images of a scene. If an estimator $f_{\tilde{\V \theta}}$ is available to produce pre-estimations $\hat{\V x}_1$ and $\hat{\V x}_2$ of the log-reflectivity images corresponding to the two speckle-corrupted log intensities $\V y_1$ and $\V y_2$, changes in the second image $\V y_2$ can be partially compensated by forming the image: $\V y_2-\hat{\V x}_2+\hat{\V x}_1$ which more closely resembles image $\V y_1$.

\begin{figure*}[ht]
	\centering
	\includegraphics[width=\textwidth]{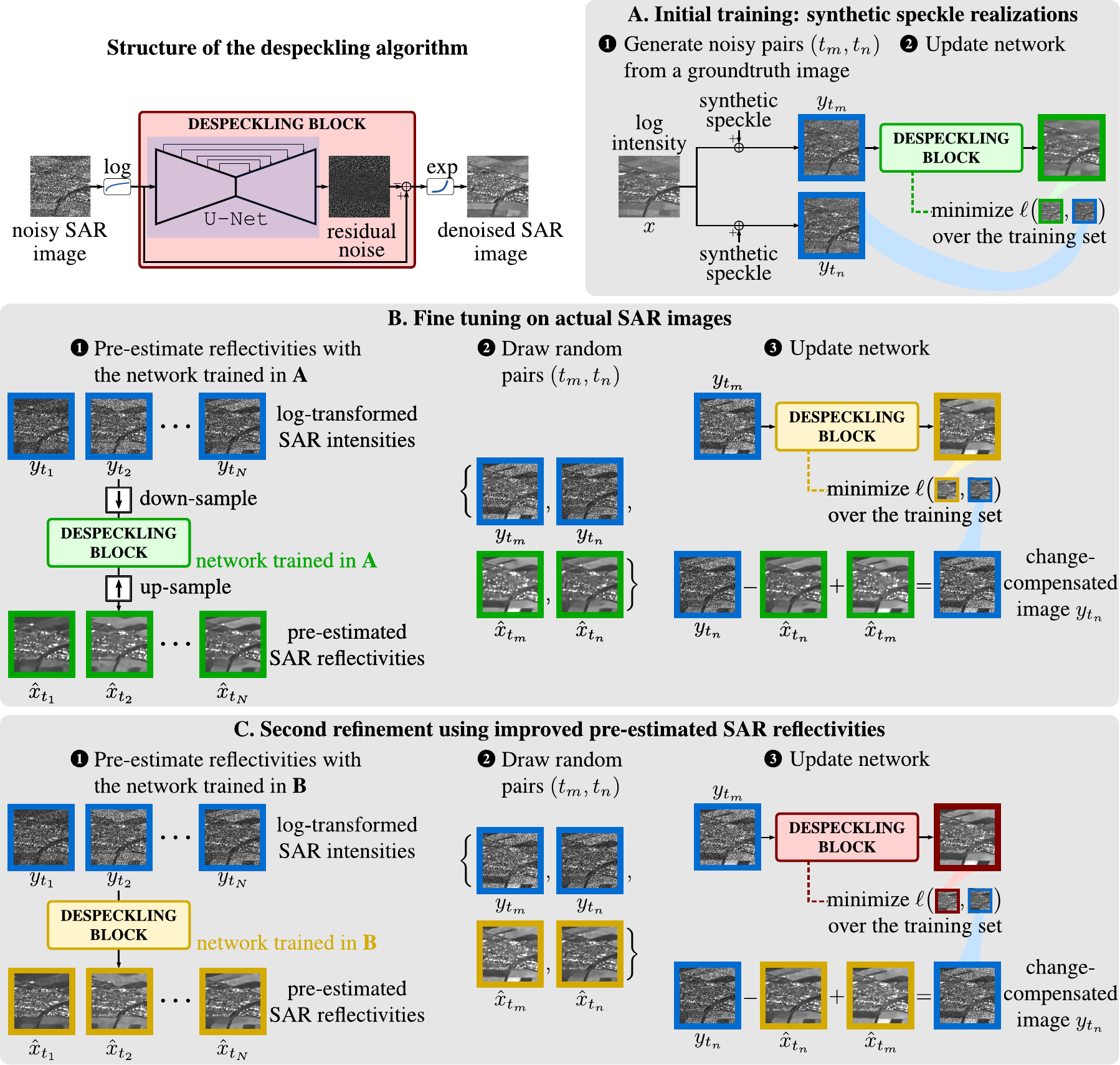}
	\caption{The proposed despeckling algorithm SAR2SAR is based on a U-Net trained in a residual learning fashion to restore log-transformed SAR images. The training is done in 3 steps: (A) on images with simulated speckle, (B) on pairs of actual SAR images, with changes compensated for based on the network trained in (A), (C) a final refinement of the network is performed with a better compensation for changes.}
	\label{fig:ppl}
\end{figure*}

The proposed SAR restoration method, named \texttt{SAR2SAR} since it extends on the original idea of \texttt{noise2noise}, considers several SAR time series, each accurately co-registered, and performs the training of a despeckling network using both the idea of the self-supervised loss of equation \ref{eq:lossSAR} and of change compensation. Figure \ref{fig:ppl} summarizes the principle of the method: the restoration is performed in the log-domain by a deep network. Since the change compensation requires the availability of pre-estimated reflectivities, the training of the network is performed in 3 steps: (A) first on images with synthetically generated speckle, (B) then on pairs of images extracted randomly from a time-series, the second image being compensated for changes based on reflectivites estimated with the network trained in (A), (C) finally a refinement step is performed where the network weights in (B) are used to obtain a better compensation for changes. As part (A) is not, in the strict sense of the term, self-supervised, we refer to \texttt{SAR2SAR} as a semi-supervised algorithm.

\section{Experiments}
\label{sec:expe}
In our set of experiments, our model is the U-Net \cite{ronneberger2015u} described in \cite{lehtinen2018noise2noise}, trained in a residual fashion \cite{he2016deep}. Images are fed to the network after a log transform. Thus, the network reproduces the noise, which is subtracted from the input image. The despeckled image is obtained as a result. 

\subsection{Synthetic speckle noise}
\begin{figure}[ht]
	\centering
	
	\begin{subfigure}{.19\textwidth}
 		\centering
 		\includegraphics[width=0.95\linewidth]{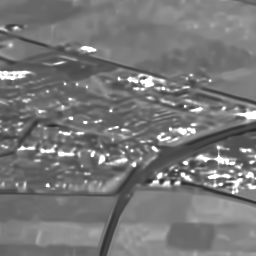}
 		\caption{Groundtruth}
 	\end{subfigure}
 	\begin{subfigure}{.19\textwidth}
 		\centering
 		\includegraphics[width=0.95\linewidth]{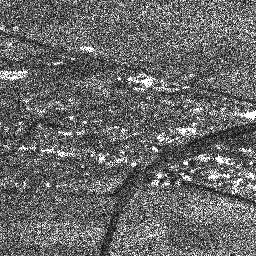}
 		\caption{Noisy}
 	\end{subfigure}%
	
    \begin{subfigure}{.19\textwidth}
 		\centering
 		\includegraphics[width=0.95\linewidth]{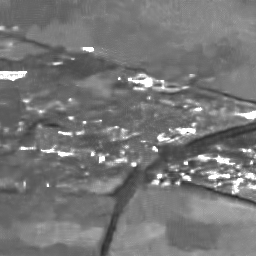}
 		\caption{SARBM3D}
 	\end{subfigure}
 	\begin{subfigure}{.19\textwidth}
 		\centering
 		\includegraphics[width=0.95\linewidth]{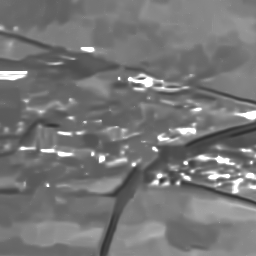}
 		\caption{MuLoG+BM3D}
 	\end{subfigure}%
 	
 	\begin{subfigure}{.19\textwidth}
 		\centering
 		\includegraphics[width=0.95\linewidth]{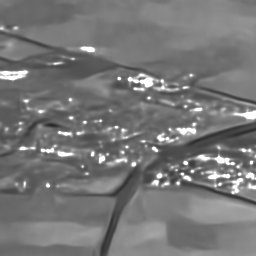}
 		\caption{SAR-CNN}
 	\end{subfigure}
 	\begin{subfigure}{.19\textwidth}
 		\centering
 		\includegraphics[width=0.95\linewidth]{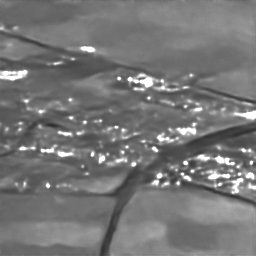}
 		\caption{noise2noise}
 	\end{subfigure}
 	\begin{subfigure}{.19\textwidth}
 		\centering
 		\includegraphics[width=0.95\linewidth]{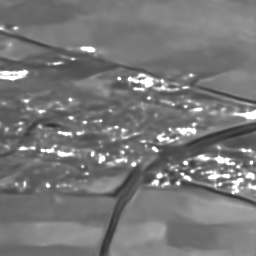}
 		\caption{SAR2SAR\textsubscript{A}}
 	\end{subfigure}
 	
	\caption{Results of state-of-the art despeckling filters on an image corrupted with synthetic 1-look uncorrelated speckle.}
	\label{fig:comparisonlely}
\end{figure}

\begin{figure}[ht]
	\centering
	
	\begin{subfigure}{.19\textwidth}
 		\centering
 		\captionsetup{justification=centering}
 		\includegraphics[width=0.95\linewidth]{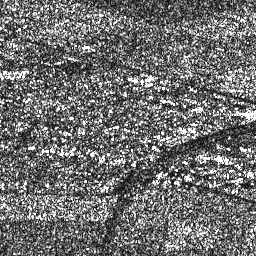}
 		\caption{Real SAR\\L=1}
 	\end{subfigure}
    \begin{subfigure}{.19\textwidth}
 		\centering
 		\captionsetup{justification=centering}
 		\includegraphics[width=0.95\linewidth]{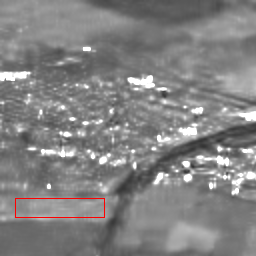}
 		\caption{SARBM3D,\\ENL = 58.00}
 	\end{subfigure}%
 	
 	\begin{subfigure}{.19\textwidth}
 		\centering
 		\captionsetup{justification=centering}
 		\includegraphics[width=0.95\linewidth]{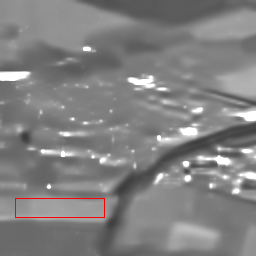}
 		\caption{MuLoG+BM3D,\\ENL = 96.20}
 	\end{subfigure}
 	\begin{subfigure}{.19\textwidth}
 		\centering
 		\captionsetup{justification=centering}
 		\includegraphics[width=0.95\linewidth]{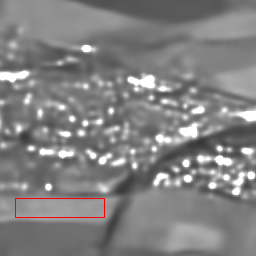}
 		\caption{SAR-CNN,\\ENL = 136.74}
 	\end{subfigure}%
 	
 	\begin{subfigure}{.19\textwidth}
 		\centering
 		\captionsetup{justification=centering}
 		\includegraphics[width=0.95\linewidth]{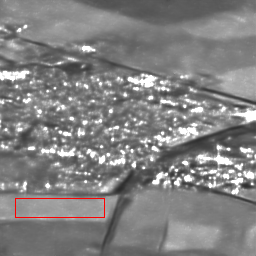}
 		\caption{SAR2SAR\textsubscript{B},\\ENL = 154.42}
 	\end{subfigure}
 	\begin{subfigure}{.19\textwidth}
 		\centering
 		\captionsetup{justification=centering}
 		\includegraphics[width=0.95\linewidth]{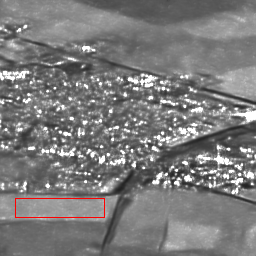}
 		\caption{SAR2SAR\textsubscript{C},\\ENL = 105.33}
 	\end{subfigure}
 	
	\caption{Results of state-of-the art despeckling filters on a real Sentinel-1 SAR image. SAR2SAR\textsubscript{B} refers to the first fine-tuning step, when images with a poorer resolution are used to compensate for changes.}
	\label{fig:comparison_rambouillet}
\end{figure}

One of the main issues when using deep learning algorithms on SAR images is the scarcity of training data. To achieve the desired level of generalization and given that the application of the presented algorithm to time-series needs an accurate adaptation, training is initially carried out on images corrupted with synthetic speckle noise. At each iteration, two independent speckle realizations (following the model described in section \ref{sec:specklemodel}) are used to create two noisy images, one being the input image and the other one to compute the loss. The images are divided into patches of $256\times256$ pixels, with a stride of 32. 3014 batches of 4 images compose our training set. The network has been trained for 30 epochs using the Adam optimizer, with a learning rate set as 0.001 and decreased by a factor of 10 after the first 5 epochs and by a factor of 100 after the first 10 epochs. The loss function of our \texttt{SAR2SAR} method is adapted to the distribution of SAR images using equation \ref{eq:lossSAR}. 

Creating synthetic images from noise-free references, moreover, allows a more reliable evaluation. Results of several despeckling filters are presented in table \ref{table:comparison_psnr}. The PSNR values on the \texttt{SAR2SAR} are not only comparable to those obtained with SAR-CNN \cite{dalsasso2020sar}, but are superior to when $\ell_2$ is adopted (with the proper debiasing step, as discussed in section \ref{sec:sar2sar_losses}), justifying the adaptation of the loss. Results on image Lely are displayed in figure \ref{fig:comparisonlely}. While the use of \texttt{SAR2SAR} is motivated by its direct application on real SAR images, even on images corrupted with synthetic speckle noise it achieves state-of-the-art results. 

\begin{table*}[htbp]
	\centering
	\begin{tabular}{l c c c c c c c}
	\toprule
	Images      & Noisy            & SAR-BM3D          & NL-SAR            & MuLoG+BM3D       & SAR-CNN & noise2noise  & \textbf{SAR2SAR\textsubscript{A}}\\
    \midrule
	Marais 1    & 10.05$\pm$0.0141 & 23.56$\pm$0.1335 & 21.71$\pm$0.1258 & 23.46$\pm$0.0794 & 24.65$\pm$0.0860 & 25.31$\pm$0.1077 & \textbf{25.73}$\pm$0.1251\\
	Limagne     & 10.87$\pm$0.0469 & 21.47$\pm$0.3087 & 20.25$\pm$0.1958 & 21.47$\pm$0.2177 & 22.65$\pm$0.2914 & 24.08$\pm$0.1099& \textbf{24.45}$\pm$0.1190\\
	Saclay      & 15.57$\pm$0.1342 & 21.49$\pm$0.3679 & 20.40$\pm$0.2696 & 21.67$\pm$0.2445 & 23.47$\pm$0.2276 & 23.50$\pm$0.3852& \textbf{23.60}$\pm$0.4368\\
	Lely        & 11.45$\pm$0.0048 & 21.66$\pm$0.4452 & 20.54$\pm$0.3303 & 22.25$\pm$0.4365 & \textbf{23.79}$\pm$0.4908 & 23.25$\pm$0.3671& 23.67$\pm$0.5415\\
	Rambouillet &  8.81$\pm$0.0693 & 23.78$\pm$0.1977 & 22.28$\pm$0.1132 & 23.88$\pm$0.1694 & \textbf{24.73}$\pm$0.0798 & 23.73$\pm$0.3882& 24.16$\pm$0.3846\\
	Risoul      & 17.59$\pm$0.0361 & 29.98$\pm$0.2638 & 28.69$\pm$0.2011 & 30.99$\pm$0.3760 & \textbf{31.69}$\pm$0.2830 & 29.93$\pm$0.2164& 30.68$\pm$0.2302\\
	Marais 2    &  9.70$\pm$0.0927 & 20.31$\pm$0.7833 & 20.07$\pm$0.7553 & 21.59$\pm$0.7573 & 23.36$\pm$0.8068 & 26.15$\pm$0.2114& \textbf{26.63}$\pm$0.2154\\
	\midrule
	Average    &  12.00 & 23.17 & 21.99 & 23.62 & 24.91 & 25.13 & \textbf{25.56}\\
    \bottomrule		
	\end{tabular}
    \caption{Comparison of denoising quality in terms of PSNR on
      amplitude images. For each ground truth image, 20 noisy
      instances are generated. 1$\sigma$ confidence intervals are
      given. Per-method averages are given at the bottom.}
%The average PSNR and its variance are separated by semicolon.}
	\label{table:comparison_psnr}
\end{table*}

\subsection{Real SAR images}

%After this initial training stage, a fine-tuning step is performed using SAR time series. The network thus adapts to the statistics of real SAR images and properly learns how to deal with spatial correlation. To compensate for changes, the results of the SAR-CNN presented in [our paper] are employed. Training proceeds for 20 more epochs with a learning rate decreased by a factor of 100 w.r.t. the initial value. 5 time series of 53 (Limagne), 45 (Marais 1), 45 (Marais 2), 69 (Rambouillet) and 25 (Lely) dates compose the training set. 2896 image patches are organized into 724 batches of 4 patches each. 

%In our experiments, the filter we use to compensate for changes between pairs of images is SAR-CNN [our paper]. As the method requires a subsampling step prior to denoising, these images are not at full resolution, having an impact on the results produced by SAR2SAR. To overcome this issue, an iterative process has been studied: every 10 epochs the a priori information given to the network is updated with the results of SAR2SAR itself. Given that, asymptotically, the function learned by the network tends to the identity, we found experimentally that one iteration is a good compromise between speckle reduction and improvement of the resolution (see table \ref{table:iterations}). Results are shown in figure \ref{fig:comparison_rambouillet}.

To fine-tune the network on real images, the SAR time series composing the training need to be denoised to generate the images used to compensate for changes. An estimation can be obtained by using the network trained on synthetic speckle, subsampling the images to reduce the effect of the correlation \cite{dalsasso2020handle}. Training proceeds for 20 more epochs with a learning rate decreased by a factor of 100 w.r.t. the initial value. Five time series of respectively 53 (Limagne), 45 (Marais 1), 45 (Marais 2), 69 (Rambouillet) and 25 (Lely) dates compose the training set. 2896 image patches are organized into 724 batches of 4 patches each. Learning is thereby transferred to correlated speckle.

\begin{table*}[htbp]
	\centering
	\begin{tabular}{l c c c c c c c c }
	\toprule
	%\text{\#} of iterations      & SAR2SAR\textsubscript{A} & SAR2SAR\textsubscript{B} & 1 & 2 & 3 & 4 & 5\\
	\multirow{3}{*}{\bf restoration method} & \multirow{3}{*}{\bf SAR2SAR\textsubscript{A}} & \multirow{3}{*}{\bf SAR2SAR\textsubscript{B}} & \multicolumn{5}{c}{\bf SAR2SAR\textsubscript{C}}\\[.5ex]
	\cline{4-8}\\[-1.5ex]
	& & & iter 1 & iter 2 & iter 3 & iter 4 & iter 5\\
    \midrule
	\bf Wasserstein distance    & 0.147 & 0.063 & 0.086 & 0.109 & 0.128 & 0.144 & 0.158 \\
    \bottomrule		
	\end{tabular}
    \caption{At each step of the iterative process, the Wasserstein distance \cite{vallender1974calculation} is computed to measure the distance between the theoretical speckle distribution of equation (\ref{eq:gamma_distribution}) and the empirical distribution of the residual speckle in intensity format (the lower the better).}
    %(\textit{i.e.} the ratio between the input noisy image and the denoised estimate).}
	\label{table:iterations}
\end{table*}

\begin{figure}[t]
	\centering
	\rotatebox{90}{Limagne} 
	\begin{subfigure}{.19\textwidth}
 		\centering
 		\includegraphics[width=0.95\linewidth]{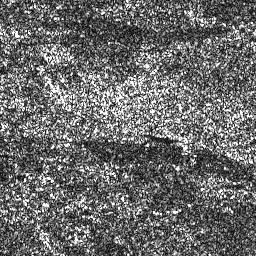}
 		\caption{}
 	\end{subfigure}
 	\begin{subfigure}{.19\textwidth}
 		\centering
 		\includegraphics[width=0.95\linewidth]{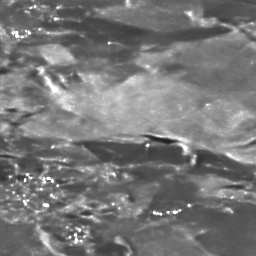}
 		\caption{}
 	\end{subfigure}%
	
	\rotatebox{90}{Marais 1} 
    \begin{subfigure}{.19\textwidth}
 		\centering
 		\includegraphics[width=0.95\linewidth]{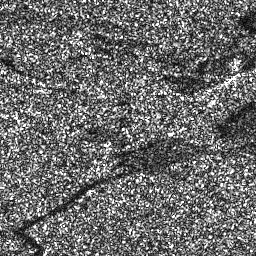}
 		\caption{}
 	\end{subfigure}
 	\begin{subfigure}{.19\textwidth}
 		\centering
 		\includegraphics[width=0.95\linewidth]{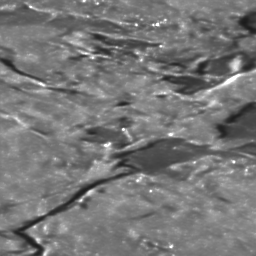}
 		\caption{}
 
 	\end{subfigure}%
 	
 	\rotatebox{90}{Rambouillet} 
 	\begin{subfigure}{.19\textwidth}
 		\centering
 		\includegraphics[width=0.95\linewidth]{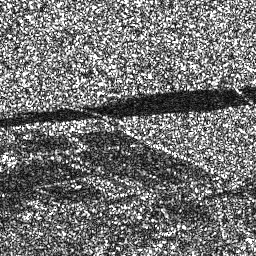}
 		\caption{}
 	\end{subfigure}
 	\begin{subfigure}{.19\textwidth}
 		\centering
 		\includegraphics[width=0.95\linewidth]{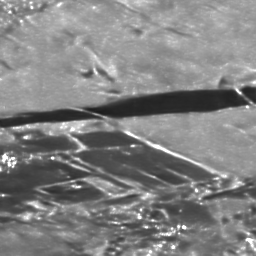}
 		\caption{}
 	\end{subfigure}%
 	
 	\rotatebox{90}{Madrid, Iowa} 
 	\begin{subfigure}{.19\textwidth}
 		\centering
 		\includegraphics[width=0.95\linewidth]{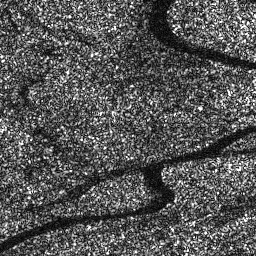}
 		\caption{}
 	\end{subfigure}
 	\begin{subfigure}{.19\textwidth}
 		\centering
 		\includegraphics[width=0.95\linewidth]{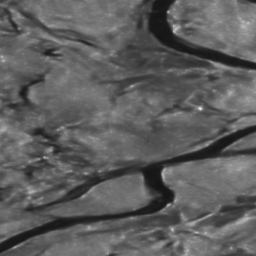}
 		\caption{}
 	\end{subfigure}%
 	
 	\rotatebox{90}{Niradpur} 
 	\begin{subfigure}{.19\textwidth}
 		\centering
 		\includegraphics[width=0.95\linewidth]{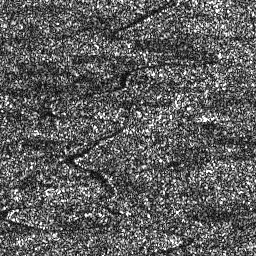}
 		\caption{}
 	\end{subfigure}
 	\begin{subfigure}{.19\textwidth}
 		\centering
 		\includegraphics[width=0.95\linewidth]{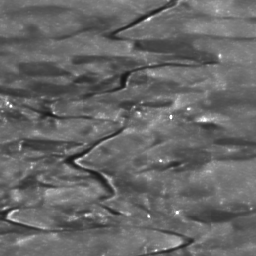}
 		\caption{}
 	\end{subfigure}%

	\caption{Restoration of Sentinel-1 images with \texttt{SAR2SAR}. Real single-look SAR images are shown on the left side, with the corresponding despeckled result on the right side.}
	\label{fig:additional_results}
\end{figure}

As the compensation images used at this step required a subsampling operation, 
they have a poor resolution that impact the results produced by 
\texttt{SAR2SAR}. To overcome this issue, an iterative process has been 
studied: every 10 epochs the compensation images given to the network are 
updated with the results of \texttt{SAR2SAR} itself. Given that, 
asymptotically, the function learned by the network tends to the identity, we 
found experimentally that one iteration is a good compromise between speckle 
reduction and improvement of the resolution (see table 
\ref{table:iterations}). Results are shown in figure 
\ref{fig:comparison_rambouillet}. In the absence of reference images, standard 
image quality metrics cannot be computed. A quantitative evaluation can be 
performed by estimating the equivalent number of looks (ENL), defined as 
$\text{ENL} = 
\mathbb{E}[\hat{\reflectivity}]^2/\text{Var}[\hat{\reflectivity}]$, on a 
manually selected homogeneous area. The higher this metric, the stronger the 
speckle reduction. Since the ENL favors filters that perform an 
over-smoothing, it cannot be used as a sole indicator for the quality of noise 
reduction. Together with visual inspection, it seems fair to say that 
\texttt{SAR2SAR} leads to the best restoration quality. Visual inspection of images
shown in figure \ref{fig:additional_results} demonstrate the effectiveness of the algorithm
in different situations: textures in the forest area are not over-smoothed (in particular Fig. \ref{fig:additional_results}.d) and fine details are well reconstructed (see the narrow river of Fig.\ref{fig:additional_results}.j), without leaving residuals of speckle noise in the filtered images.

\section{Ablation study}
\begin{figure*}
    \centering
    \includegraphics{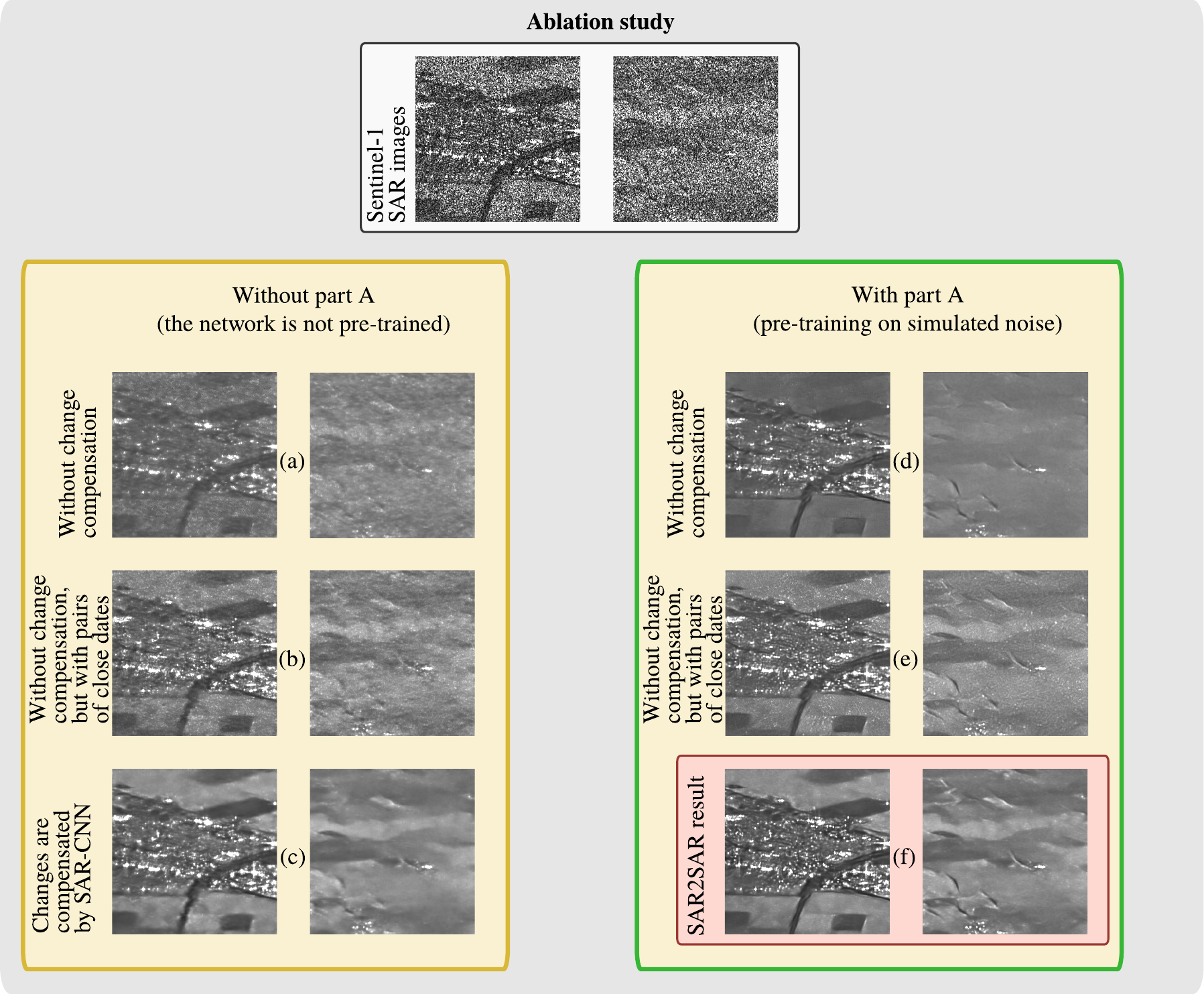}
    \caption{Ablation study of \texttt{SAR2SAR}, showing the restoration of two Sentinel-1 images. In the left column, a self-supervised training conducted only on real images is considered. On the right column, the network is initially trained on images with synthetic speckle noise. Different strategies are considered, showing the importance of change compensation and the relevance of the proposed semi-supervised approach.}
    \label{fig:ablation}
\end{figure*}

To demonstrate the impact of the pre-training step with synthetically 
generated speckle noise and the importance of change compensation, an ablation 
study is conducted and results on two Sentinel-1 image patches are shown in 
Figure \ref{fig:ablation}. 

The use of a change compensation mechanism plays a key role in the effective 
exploitation of multi-temporal SAR images.
%The use of information to compensate for changes is the key step that allows 
%exploitation of SAR stacks. 
%From our experiments, it turned out that t
Training directly on image pairs without compensating for change leads to poor 
results with notable bias in some areas (Fig. \ref{fig:ablation}.a and Fig. 
\ref{fig:ablation}.d). Even when only the closest date is selected to minimize 
changes, the lack of change compensation is penalizing and additional 
artifacts also occur due to the non-negligible temporal correlation of speckle 
(Fig. \ref{fig:ablation}.b and Fig. \ref{fig:ablation}.e).

When the network is not pre-trained, i.e., there is no step A with synthetic 
speckle and only pairs of actual SAR images are used, the restoration 
performance worsens (left part of Fig. \ref{fig:ablation}, in particular Fig. 
\ref{fig:ablation}.c). A reason for this 
drop in image quality may be that the learning is guided uniquely by the 
change-compensated image. Since the method to compensate the changes is not 
perfect, the images that drive the estimation of the network weights are of 
lesser quality. This could possibly be mitigated by considering much larger 
training sets.
%
%in a scenario under which temporal stability can be assumed was not leading 
%to 
%such good results. 
%To do this, given an image at date $t$ as input, the reference image has to be picked at $t\pm1$. Thus, the hypothesis of "no changes" does not hold in a real situation. 
%To do this, given an image at date $t$ as input, the closest image in time 
%was chosen to do the training. But even in this condition, the no change 
%hypothesis did not always hold, giving mitigated results.
%
%When the network is trained in a fully self-supervised framework 
%(\textit{i.e.} by only using real SAR images), a filter such as SAR-CNN 
%\cite{dalsasso2020sar} can be employed to pre-estimate SAR reflectivities. As 
%the pre-estimates have undergone a subsampling step, changes are not 
%perfectly 
%compensated, preventing a full training in a self-supervised framework. 
%
Warm-starting the network using simulated data allows to create a virtually 
unlimited number of noisy image pairs of the same area. When trained this way, 
the model is competitive with other state-of-the-art despeckling filters (see 
Table \ref{table:comparison_psnr}) and only needs a slight adjustment to
gain robustness to the spatial correlations of the speckle. It can then be 
used to produce pre-estimates of the SAR reflectivities to compensate for 
temporal changes, and refined on real images with our two-steps self-supervised 
strategy, where step (C) removes the initial limitations due to the 
down-sampling procedure used to decorrelate the speckle.
%due to the initial subsampling applied when compensating for changes with 
%images restored with a despeckling .

\section{Discussion} \label{sec:discussion}
Single-look SAR images are difficult to denoise due to the strong spatial correlation. The proposed SAR2SAR algorithm learns the statistics of real speckle noise directly from the data, by devising a self-supervised algorithm leveraging deep learning and multi-temporal stacks of images. 

While providing state-of-the-art results on images with synthetic speckle noise, it is on real single-look images that SAR2SAR shows a clear improvement over existing despeckling algorithms. The methods developed under fully developed speckle model assumptions, indeed, need a careful adaptation in order to properly deal with correlated data \cite{dalsasso2020handle}. If a subsampling step is applied, images with a poorer resolution are produced. A more careful pre-processing, however, needs knowledge of the sensor's parameters and adds a computational burden. SAR2SAR learns the speckle model directly from the data, making it readily applicable on real images. The advantage of deep learning algorithms, moreover, is that they are computationally fast once they are trained.

%Training from scratch on real data has also been considered. In this case, the changes compensation rely on an external denoiser. In our experiments, SAR-CNN [our paper] has been used. However, the degree of uncertainty given by the estimations of SAR-CNN, leading to a non-perfect compensation, and the limited amount of data prevented to obtain high-quality results and strong generalization capabilities. An initial training phase using a synthetic dataset is thus preferable. Moreover, it allows an estimation of the compensation images used when fine-tuning on real data.

The general formulation of SAR2SAR suggests that it can be extended to Ground Range Detected (GRD) images and to any sensor (e.g. TerraSAR-X), once the training data are collected. The weights of the trained model are released along with this article\footnote{\url{https://gitlab.telecom-paris.fr/RING/SAR2SAR}}, to allow testing of our method and to foster research on SAR image denoising. Additional results are also provided in our Git repository.
\ifCLASSOPTIONcaptionsoff
  \newpage
\fi

\bibliographystyle{IEEEtran}
\bibliography{ref}
\end{document}